\documentclass[11pt]{article}

\usepackage[utf8]{inputenc}  
\usepackage[english]{babel}
\usepackage{amsmath}
\usepackage{amsfonts}
\usepackage{amssymb}
\usepackage{bm}
\usepackage{graphicx}
\usepackage{indentfirst}
\usepackage{url}
\usepackage[hang, nooneline]{subfigure}

\usepackage{geometry}
\begin{document}

\title{Encapsulated formulation of the Selective Frequency Damping method} 

\author{Bastien E. Jordi, Colin J. Cotter, Spencer J. Sherwin} 

%

\date{\today}

\maketitle

\begin{abstract}
We present an alternative ``encapsulated'' formulation of the Selective Frequency Damping method for finding unstable equilibria of
dynamical systems, which is particularly useful when analysing the
stability of fluid flows. The formulation makes use of splitting methods, which means that it can be wrapped around an existing
time-stepping code as a ``black box''. The method is first applied to a
scalar problem in order to analyse its stability and highlight the
roles of the control coefficient $\chi$ and the filter width $\Delta$
in the convergence (or not) towards the steady-state. Then the
steady-state of the incompressible flow past a two-dimensional
cylinder at $Re=100$, obtained with a code which implements the
spectral/\textit{hp} element method, is presented.
\end{abstract}

\section{Introduction}

A deep understanding of the underlying physics of boundary layer
separation behaviour is crucial when considering the problem of
controlling detached flows. Flow instabilities may be the cause of
this separation. A stability analysis of a flow can help to predict
when these instabilities arise, and can be used to design flow
controllers. To perform a stability analysis, it is necessary to
construct the ``base flow'' around which the system will be
linearised. This leads to the challenge of obtaining a steady state
solution of the Navier-Stokes equations. There are alternatives to
this, namely to compute the time average of an unsteady solution, or
to obtain a solution of the Reynolds averaged Navier-Stokes (RANS) equations
to use as the base flow. However, these solutions often produce
stability properties which are not relevant to the physical solution
\cite{BarkleyMeanVsSteadyCylinderFlow}. Hence, we are interested in solver algorithms to obtain genuine steady-state solutions of the
Navier-Stokes equations. One possible approach is to use Newton's
method, however it must be combined with a continuation method since a
good initial guess of the solution is required to ensure convergence.
For challenging flow problems at high Reynolds number, it may be the
case that many complicated bifurcations at various Reynolds numbers
must be crossed before finding the required solution. As an
alternative, $\mbox{\AA}$kervik \textit{et al.}  \cite{Steady_NS_SFD}
presented a modification of the time-dependent dynamical system,
called Selective Frequency Damping (SFD), which tries to reach the
steady-state of an unsteady system by damping unstable temporal
frequencies. Since it is easy to implement into an existing code, and
does not need a good initial guess, this method appeared to be an
efficient alternative to classical Newton's methods. The SFD method
has been successfully applied to find steady-solutions of the
Navier-Stokes equations, which were then used as a base flow to study stability
properties of flows such as the wake of a sphere
\cite{SFDusedFor3DshpereFlow}, a jet in a crossflow
\cite{SFDforJetFlow} or a cavity flow
\cite{SFDoptimalGrowth}. However Jones and Sandberg \cite{SFDfailing1}
failed to find the steady-state of the compressible flow around a
NACA-0012 airfoil at $Re=1 \times 10^6$ using the SFD method. It was
assumed there that the method was not able to suppress the
instabilities present at this Reynolds number without requiring a
large damping coefficient and hence impractically long
time-integration to converge. Vyazmina \cite{SFDfailing2} studied
swirling flows and it was noticed that the SFD method did not converge
towards the steady-state when the problem considered had real unstable
eigenvalues.

In this paper we present a time discrete formulation of the SFD method
which is implemented as a wrapper around an existing "black box"
unsteady solver. In section \ref{SectionProblemFormulation} we first
recall the original form of the SFD method and then show that with the
splitting methods framework the method can be reformulated. This
alternative formulation encapsulates the existing solver. This scheme
is applied to a simple one-dimensional problem in section
\ref{1DPrb}. The convergence properties of this formulation are
studied in order to provide information about the influence of the
filter width and the control coefficient on the stability of the
method. Then the results obtained by the application of the
encapsulated SFD method to a high-order incompressible Navier-Stokes
solver are presented in section
\ref{SectionNumericalSimulation}. Finally, we introduce the idea that
isolating the most unstable eigenmode of a flow problem and treating
it as a one-dimensional problem can give sufficient information in
order to ensure the convergence of the SFD problem applied to a
Navier-Stokes solver.

\section{Problem formulation}
\label{SectionProblemFormulation}

We first recall the basis of the SFD method as
it was originally introduced. Then we present the discretized
encapsulated formulation.

With appropriate initial and boundary conditions, any system can be
written
\begin{equation}
\dot{q}=f(q),
\label{system}
\end{equation}
where $q$ represents the problem unknown(s), the dot represents the time derivative and $f$ is an operator (which
can be nonlinear). The
steady-state $q_s$ of this problem is reached when
$\dot{q_s}=f(q_s)=0$.

The main idea of the SFD method is to
introduce a linear forcing term on the right-hand side of
(\ref{system}). This term must contain a control coefficient and a
target towards which the solution will be driven to. A new problem
formulation is then defined such as
\begin{equation}
\dot{q}=f(q)-\chi (q - q_s),
\label{SFD}
\end{equation}
where $\chi$ is the control coefficient and $q_s$ is the target
steady-state. This stabilization technique is called proportional
feedback control and is commonly used in control theory
\cite{kim2007linear}. When $q_s$ is a genuine steady-state,
\emph{i.e.}, $f(q_s)=0$, the steady solution of (\ref{SFD}) is clearly
also the steady solution of (\ref{system}). However, in practice,
especially for real flow problems, the steady-state is generally not
known \textit{a priori}. The SFD method addresses this by replacing
$q_s$ by a low-pass filtered version of $q$, denoted $\bar{q}$. By damping the
most dangerous frequencies, the corresponding instabilities are
extinguished \cite{Steady_NS_SFD}. This idea was originally introduced
by Pruett \textit{et al.} \cite{Pruett1, Pruett2} in their work on a
temporal filtered model developed for large-eddy simulations.

The differential form of a (first order) low-pass time filter can be
defined as
\begin{equation}
\dot{\bar{q}}=\frac{q-\bar{q}}{\Delta},
\label{filter}
\end{equation}
where $\bar{q}$ is the temporally filtered quantity and $\Delta$ is
the filter width. This equation can be advanced in time using any
appropriate integration scheme.

Considering (\ref{SFD}), with the new target solution $\bar{q}$, and
the filter (\ref{filter}), we obtain the system
\begin{equation}
\begin{cases}
\dot{q} = f(q)-\chi (q-\bar{q}), \\
\dot{\bar{q}} = \frac{q-\bar{q}}{\Delta}.
\label{SFD-General}
\end{cases}
\end{equation} 

This system is the continuous time formulation of the SFD method, as it
was first introduced \cite{Steady_NS_SFD}. The filtered solution
$\bar{q}$ is time varying, and the steady-state is reached when $q =
\bar{q}$. We now present a time-discrete implementation of the SFD
method which allows us to wrap code around an existing time-stepping
scheme for equation \ref{system}. This notion can be linked to
Tuckerman's work \cite{TuckermanBifurcationAnalysisForTimeSteppers} on
the adaptation of time-stepping codes to carry out efficient
bifurcation analysis.

System (\ref{SFD-General}) can be discretised within the framework of
sequential operator-splitting methods \cite{farago2005splitting}. The
system is divided into two smaller subsystems which are solved
separately using different numerical schemes. The first subproblem
(which can be nonlinear) is simply (\ref{system}). We introduce the
function $\Phi$ such that the numerical (or exact) solution of
(\ref{system}) at the step $(n+1)$ is given by
\begin{equation}
q^{n+1} = \Phi (q^n).
\label{Subproblem1}
\end{equation}

The second subproblem is linear and represents the actions of the
feedback control and the low-pass time filter. It can be formulated
\begin{equation}
\begin{cases}
\dot{q}=-\chi (q-\bar{q}) \\
\dot{\bar{q}}=\frac{q-\bar{q}}{\Delta}
\end{cases}
\Longleftrightarrow
\begin{pmatrix}
\dot{q} \\ 
\dot{\bar{q}}
\end{pmatrix}
= \begin{pmatrix}
- \chi I &  \chi I \\ 
I/\Delta & -I/\Delta
\end{pmatrix} \begin{pmatrix}
q \\ 
\bar{q}
\end{pmatrix},
\label{Subproblem2}
\end{equation}
where $I$ is the identity matrix. The linear operator defined by
(\ref{Subproblem2}) will be denoted $\mathcal{L}$. This equation can
be solved exactly on $[t^n,~ t^n + \Delta t]$ and the solution is
given by
\begin{equation}
\begin{pmatrix}
q(t^{n+1}) \\ 
\bar{q}(t^{n+1})
\end{pmatrix}
=
e^{\mathcal{L} \Delta t}
\begin{pmatrix}
q(t^n) \\ 
\bar{q}(t^{n})
\end{pmatrix},
\label{ExactSolSubproblem2}
\end{equation}
where the expanded expression of $e^{\mathcal{L} \Delta t}$ is
\begin{equation}
e^{\mathcal{L} \Delta t} =  \frac{1}{1 + \chi \Delta} \times \begin{pmatrix}
I + \chi \Delta I e^{-(\chi + \frac{1}{\Delta})\Delta t} & \chi \Delta I [1 - e^{-(\chi + \frac{1}{\Delta})\Delta t}]\\ 
I - I e^{-(\chi + \frac{1}{\Delta})\Delta t} & \chi \Delta I + I e^{-(\chi + \frac{1}{\Delta})\Delta t}
\end{pmatrix}.
\label{ExpLinearOperator_Developed}
\end{equation}

In the construction of a splitting method, the final solution of one
subproblem is used as initial condition of the other one. As
(\ref{Subproblem1}) does not affect $\bar{q}$, the discrete
formulation of (\ref{SFD-General}) using a first order splitting
method is given by
\begin{equation}
\begin{pmatrix}
q^{n+1} \\ 
\bar{q}^{n+1}
\end{pmatrix}
=
e^{\mathcal{L} \Delta t}
\begin{pmatrix}
\Phi (q^n) \\ 
\bar{q}^{n}
\end{pmatrix},
\label{SplittingSFD}
\end{equation}
where $\Delta t$ is the time step used within the solver $\Phi$. We
call this scheme ``encapsulated'' since $\Phi$ is not modified but
simply used as an input of the linear solver
(\ref{ExactSolSubproblem2}).  Hence $\Phi$ can be treated as a ``black
box''. Codes which solve the Navier-Stokes equations are usually very
complicated. If the original problem (\ref{system}) is a flow problem,
implementing an efficient steady-state solver with minimum programming
effort can be highly valuable. To implement (\ref{SplittingSFD}) in an
existing code, the only work required is to create an auxiliary
variable $q^*$ which takes the value of the outcome of
(\ref{Subproblem1}). Then the linear operator $e^{\mathcal{L} \Delta
  t}$ (which is constant through time) simply has to be applied to the
vector $(q^*, ~\bar{q}^{n})^T$. Note that a second order Strang
splitting method can also be used to solve (\ref{SFD-General}). For
clarity we only present the first order scheme here.

This method does not converge to a steady-state for arbitrary control
coefficient $\chi$ and filter width $\Delta$. If $\Phi$ is a linear
map, the convergence of (\ref{SplittingSFD}) towards the steady-state
of (\ref{system}) is guaranteed if all the eigenvalue magnitudes of
this system are strictly smaller than one. Such a system is said to be
(linearly) stable. As (\ref{SplittingSFD}) depends on $\chi$ and
$\Delta$, these parameters play a key role in the stability of the SFD
method. In the next section, we analyse this role, using a
one-dimensional model.

\section{Scalar problem}
\label{1DPrb}

In this section a simple one-dimensional problem is studied in order
to analyse the influence of $\chi$ and $\Delta$ on the stability of
the SFD method. A clear understanding of their role should help users
of the SFD method to choose parameters that ensure its
convergence. The scalar problem considered is
\begin{equation}
\dot{u} = \gamma u,
\label{1DEquation_continuous}
\end{equation} 
where $\gamma \in \mathbb{C}$. This equation has exact solver
\begin{equation}
u^{n+1} = \Phi_{1\text{D}}(u^n) = \alpha u^n, ~ \alpha = e^{\gamma \Delta t}.
\label{1DEquation}
\end{equation} 
For the remainder of this section we set $\Delta t = 1$.

The convergence towards the steady-state of (\ref{1DEquation})
(\textit{i. e. $u^{n+1} = u^n$}) is guaranteed if $|\alpha|<1$. We aim
to use SFD to reach the steady-state of (\ref{1DEquation}) when
$|\alpha| > 1$. Analysing this simple case will allow us to highlight
the roles of the parameters $\chi$ and $\Delta$ in the convergence (or
not) of the SFD method.

In order to write the encapsulated formulation of the SFD method
applied to (\ref{1DEquation}), we use the function
$\Phi_{1\text{D}}$ such that $\Phi_{1\text{D}}(u^n) = \alpha
u^n$. Then we introduce the operator $\mathcal{L}_{1\text{D}} \in
\mathcal{M}_2 (\mathbb{R})$ which has the same form as
(\ref{Subproblem2}) upon replacing $I$ with $1$. The application of
the encapsulated formulation of the SFD method to (\ref{1DEquation})
becomes
\begin{equation}
\begin{pmatrix}
u^{n+1} \\ 
\bar{u}^{n+1}
\end{pmatrix}
=
e^{\mathcal{L}_{1\text{D}}}
\begin{pmatrix}
\alpha u^n \\ 
\bar{u}^{n}
\end{pmatrix}
=
e^{\mathcal{L}_{1\text{D}}}
\begin{pmatrix}
\alpha & 0\\ 
0 & 1
\end{pmatrix}
\begin{pmatrix}
u^n \\ 
\bar{u}^{n}
\end{pmatrix}.
\label{1D_SplittingSFD}
\end{equation}

The eigenvalues (noted $\lambda_1$ and $\lambda_2$) of the matrix
defined by system (\ref{1D_SplittingSFD}) can then easily be
evaluated, as functions of the control coefficient $\chi$, the filter
width $\Delta$ and the complex number $\alpha$. To ensure the
stability of (\ref{1D_SplittingSFD}), we want to be able to choose
$\chi$ and $\Delta$ such that max$(|\lambda_1|, |\lambda_2|)<1$.

From these eigenvalues we obtain the following limiting behaviour
for small $\chi$ and $\Delta$:
\begin{equation}
\begin{cases}
\begin{split}
&\lim\limits_{\chi \rightarrow 0} \lambda_1 = \alpha, \\
&\lim\limits_{\chi \rightarrow 0} \lambda_2 = e^{- 1 / \Delta},
\end{split}
\end{cases} ~~~
\begin{cases}
\begin{split}
&\lim\limits_{\Delta \rightarrow 0} \lambda_1 = \alpha, \\
&\lim\limits_{\Delta \rightarrow 0} \lambda_2 = 0.
\end{split}
\end{cases}
\end{equation}

If the original problem is not converging, applying the SFD
method and choosing a small control coefficient $\chi$ will not drive the
solution towards its steady-state. We also obtain the following
large $\chi$ and $\Delta$ limits:
\begin{equation}
\begin{cases}
\begin{split}
&\lim\limits_{\chi \rightarrow + \infty} \lambda_1 = 1, \\
&\lim\limits_{\chi \rightarrow + \infty} \lambda_2 = 0,
\end{split}
\end{cases} ~~~
\begin{cases}
\begin{split}
&\lim\limits_{\Delta \rightarrow + \infty} \lambda_1 = \alpha e^{-\chi}, \\
&\lim\limits_{\Delta \rightarrow + \infty} \lambda_2 = 1.
\end{split}
\end{cases}
\end{equation}

If the control coefficient $\chi$ (or the filter width $\Delta$) is
chosen to be large, the encapsulated formulation of the SFD method is
marginally stable. The steady-state can not be reached but the
solution does not blow up. 

We now examine the stability regions of the encapsulated formulations
of the SFD method. The goal, for a given control coefficient and
filter width, is to identify for which $\alpha$ the one-dimensional
problem will converge towards its steady-state. This section is only
focused on the influence of $\chi$ and $\Delta$.

\begin{figure}
\begin{center}
\subfigure[$\Delta = 0.5$]{\label{Delta05} \includegraphics[width=8.0cm]{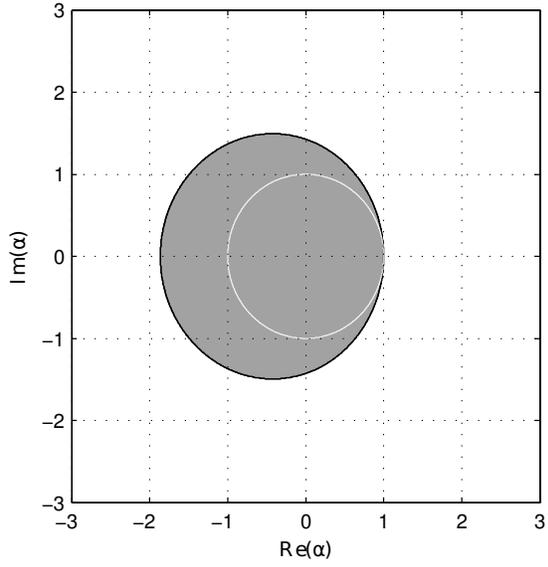}}
\subfigure[$\Delta = 2$]{\label{Delta2} \includegraphics[width=8.0cm]{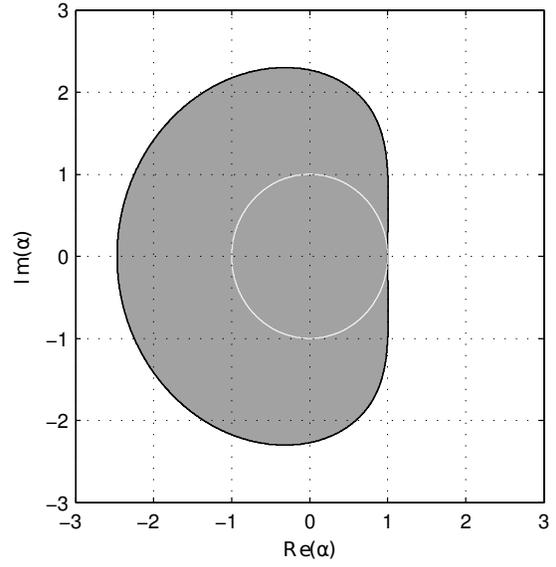}}
\subfigure[$\Delta = 20$]{\label{Delta20} \includegraphics[width=8.0cm]{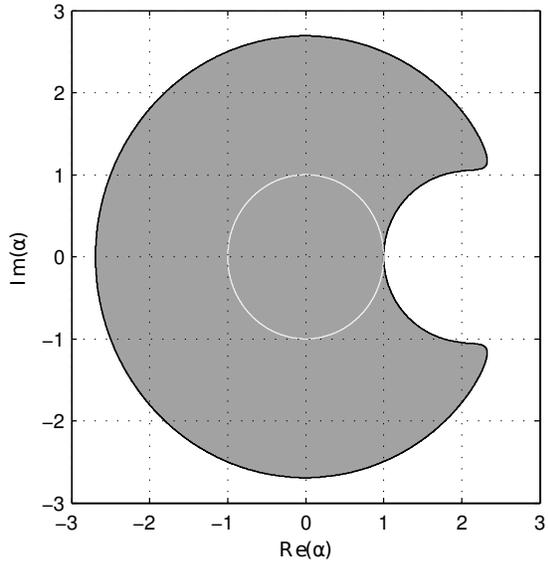}}
\subfigure[$\Delta = 10000$]{\label{Delta10000} \includegraphics[width=8.0cm]{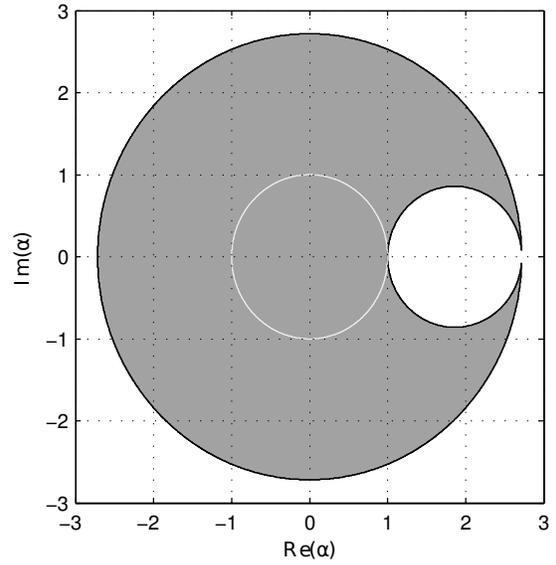}}
\caption{\label{StabRegions} Stability regions for $\chi = 1$ and
  various $\Delta$. If $\alpha$ is inside the grey area then
  (\ref{1D_SplittingSFD}) converges towards the steady-state of
  (\ref{1DEquation}). The unit circle (\textit{i. e.} the region where
  $|\alpha| = 1$) is displayed in white.}
\end{center}
\end{figure}

In the stability diagram in Figure \ref{StabRegions}, each point of the
complex plane corresponds to the value of $\alpha$. The values of
$\chi$ and $\Delta$ are fixed for every point, and the eigenvalues of
matrix (\ref{1D_SplittingSFD}) are evaluated. If both eigenvalue
magnitudes are smaller than one, then the point is coloured in
grey. Hence the grey area corresponds to the stability region of
(\ref{1D_SplittingSFD}). In other words,
Figure \ref{StabRegions}\subref{Delta05} tells us that $\chi = 1$ and
$\Delta = 0.5$ will drive (\ref{1D_SplittingSFD}) towards its
steady-state for every $\alpha$ chosen within the grey area.

We recall that (\ref{1DEquation}) was stable only if $|\alpha|<1$
(\textit{i. e.} only if $\alpha$ was situated within the unit disc,
delimited by the white circle on Figure \ref{StabRegions}). The stability
region of (\ref{1D_SplittingSFD}) expands beyond the unit disc. These
pictures allow us to visualize the fact that the SFD method stabilizes
unstable modes. This is achieved without introducing a loss of
stability elsewhere, indeed the stability region of the original
problem (\textit{i. e.} the unit disc) is inside the stability region.

On each stability diagram presented in Figure \ref{StabRegions}, we notice that if $\alpha$
is real and greater than one, it is not possible to find a couple
$\chi$ and $\Delta$ for which (\ref{1D_SplittingSFD}) is stable. Such
an $\alpha$ corresponds to a problem with a pure exponential growth of
the instability. Hence we can conclude that the stability of the SFD
method relies on the oscillatory growth of the problem studied. This
is because time averaging oscillatory growth produces a good estimate
of the equilibrium, whilst time averaging exponential growth does not.
This was reported by Vyazmina \cite{SFDfailing2}, who said that if an
unstable eigenvalue is real and positive, there is no frequency to be
damped by the SFD method.

\begin{figure}
\begin{center}
\includegraphics[width=8.0cm]{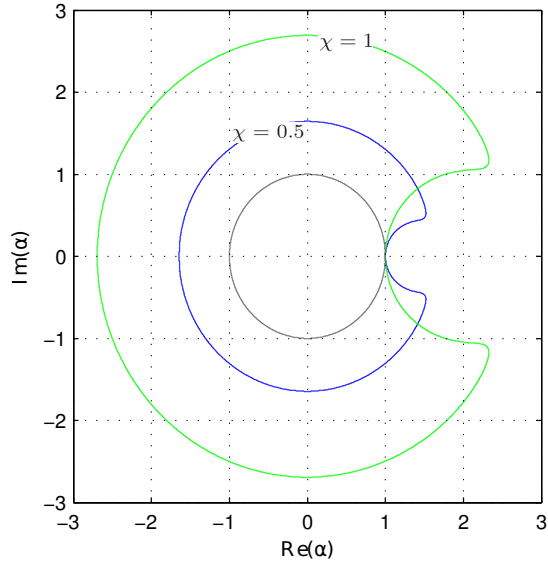} 
\caption{\label{StabContours_FixedDelta} Contours of the stability regions of
  (\ref{1D_SplittingSFD}) for $\Delta = 20$ and two different $\chi$. The
  central circle represents the boundary of the unit disc.}
\end{center}
\end{figure}

Figure \ref{StabRegions} presents stability regions for a fixed value
of the control coefficient $\chi$. With a different $\chi$, the shape
of these stability regions would have been similar but the area
covered would have been different. This behaviour is shown in Figure \ref{StabContours_FixedDelta}.

$\mbox{\AA}$kervik \textit{et al.} \cite{Steady_NS_SFD} stated that
choosing a large $\chi$ or a large $\Delta$ would make the system
evolution very slow but the SFD method would eventually converge to a
steady-state. The results presented in Figure \ref{StabContours_FixedDelta} suggest that this may not always be true. Indeed, this picture shows that there is a region where (\ref{1D_SplittingSFD}) is stable for $\chi = 0.5$ and $\Delta = 20$ but unstable for $\chi = 1$ and $\Delta = 20$. Hence increasing the control coefficient is not always an appropriate method to guarantee the convergence of the SFD method towards the steady-state.  

When choosing a large filter width, the behaviour of the SFD method is slightly different. For a given $\chi$ and a large $\Delta$, if the SFD method is not stable then it is not possible to find a smaller $\Delta$ for which the method is stable. This is illustrated by the fact that the regions presented in Figures \ref{StabRegions}\subref{Delta05}, \ref{StabRegions}\subref{Delta2} and \ref{StabRegions}\subref{Delta20} are all included within the region presented in Figure \ref{StabRegions}\subref{Delta10000}. However, choosing a very large $\Delta$ does not guarantee the stability of (\ref{1D_SplittingSFD}). If $\alpha$ is situated at the right of the unit circle, close to the real axis, increasing the filter with without acting on the control coefficient might not be enough to enable the SFD method to converge.

When $\chi = 0$ and when $\Delta$ tends to zero, the stability region of
(\ref{1D_SplittingSFD}) fits exactly within the unit circle,
confirming the outcome of the limiting behaviour analysis for the
scalar problem (for conciseness these figures are not presented here).

If a second order Strang splitting method is used to reformulate
(\ref{SFD-General}) applied to the scalar problem (\ref{1DEquation}),
the stability regions obtained are exactly the same as the ones
presented in this section. This is because the second-order splitting
can be written as a shifted first-order splitting with a pre- and
post-processing step, so both methods have the same stability regions.

\section{Numerical simulations: the cylinder flow at $Re=100$}
\label{SectionNumericalSimulation}

The encapsulated SFD method (\ref{SplittingSFD}) was implemented into
the Nektar++ spectral/\textit{hp} element framework \cite{Nektar++},
as a wrapper function. In order to find the steady state solution of the
incompressible Navier-Stokes equations, a solver which implements the
\textit{velocity-correction} scheme \cite{SencerBook} (symbolised by
the functions $\Phi$ in (\ref{SplittingSFD})) is called at each
time-step. In this section we present the numerical steady-state of
the two dimensional cylinder flow above its critical Reynolds number
$Re_c$ (\textit{i. e.} when the Reynolds number is high enough such
that the viscous forces within the flow are not dominant).

At $Re = 100$ the two-dimensional incompressible flow past a cylinder
is unstable and von Kármán vortex streets are observable. Indeed,
shedding vortices appear when $Re > Re_c \simeq 47$ and this
phenomenon remains until the end of the subcritical regime ($Re \simeq
2 \times 10^5$) \cite{CylFlowReview}. Hence the role of the SFD method
is to suppress these oscillations and drive the solution towards its
steady-state.
\begin{figure}
\begin{center}
\subfigure[Snapshot of the uncontrolled flow (unsteady and
  periodic). This behaviour is the well known von Kármán
  shedding.]{\label{UncontrolledFlow}
  \includegraphics[width=8.0cm]{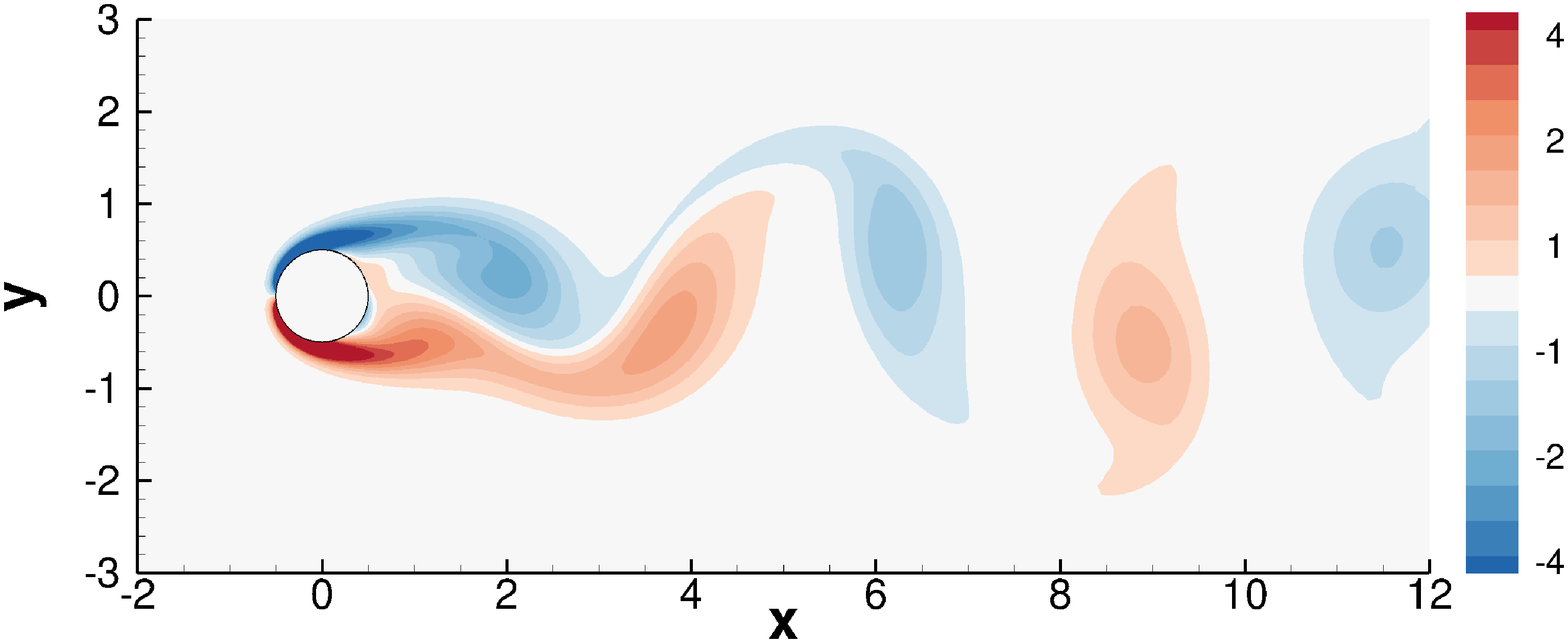}} 
  \subfigure[Unstable
  steady-state obtained by SFD. The dashed lines represent the
  separating streamlines.]{\label{SteadyFlow}
  \includegraphics[width=8.0cm]{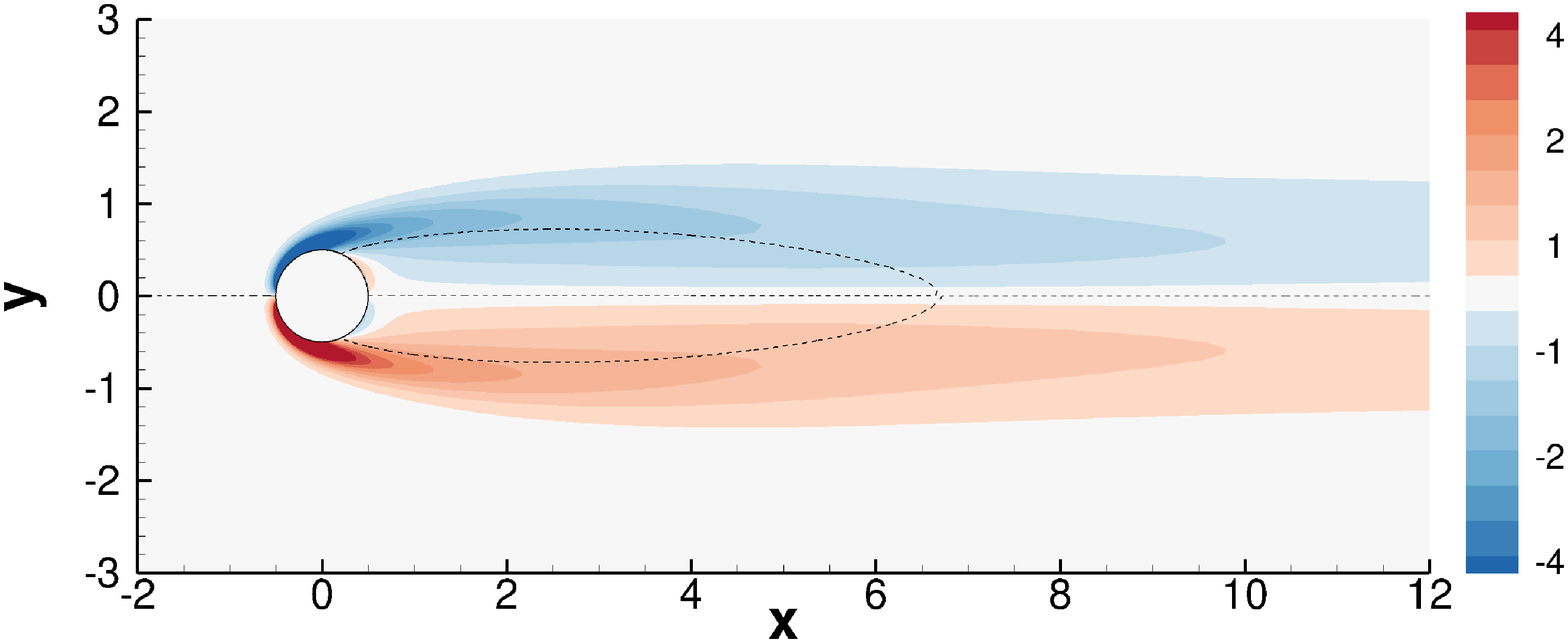}}
\caption{\label{DiplayResultX1Delta2} Vorticity ($\omega = \partial_x
  v - \partial_y u$) of the incompressible flow past a two-dimensional
  cylinder at $Re=100$. Note that only a small part of the whole
  computational domain is displayed.}
\end{center}
\end{figure}

The computational domain dimensions are $-15 \leq x \leq 45$ and $-25
\leq y \leq 25$. The cylinder is centred at the origin and its
diameter is one unit length. Close to the cylinder the mesh composed
of structured quadrilateral elements; elsewhere, the mesh is composed
of triangular elements. On the cylinder surface, no-slip boundary
conditions are imposed. Dirichlet boundary conditions $(u, ~v) = (1,
~0)$ are imposed on the left, top and bottom edges. Finally an outflow
boundary condition is set on the right edge. As the steady solution is
expected to be smooth, a high polynomial order of 11 is used. To
highlight the fact that, in contrast to Newton's methods, the SFD
method does not need a good approximation of the final solution to
converge, initial conditions such as $(u_0,~v_0) = (0,~0)$ are
chosen. The SFD parameters are initially chosen such that the control
coefficient $\chi = 1$ and the filter width $\Delta = 2$. The problem
is considered to have converged when $|| q^n - \bar{q}^n ||_{\text{inf}}
< 10^{-8}$. The steady-state has been obtained after the computation
of about 1000 time units (with the time-step $\Delta t =0.01$) and the decay of $|| q^n - \bar{q}^n
||_{\text{inf}}$ is exponential. 

Figure \ref{DiplayResultX1Delta2}\subref{UncontrolledFlow}
is simply a reminder of the behaviour of the uncontrolled incompressible cylinder
flow at $Re=100$ and the steady-state obtained by SFD is shown on
Figure \ref{DiplayResultX1Delta2}\subref{SteadyFlow}. This flow configuration is identical to
the one presented by Barkley
\cite{BarkleyMeanVsSteadyCylinderFlow}. 

A stability analysis, using an Arnoldi method, is performed with this steady-state as the "base flow". The growth rate $\sigma$ and the frequency $f$ of this flow configuration are
\begin{equation}
\sigma = 0.12978 ~\mbox{~and~}~ f = 2 \pi \times 0.11769,
\end{equation}
which correspond to the values presented by Barkley \cite{BarkleyMeanVsSteadyCylinderFlow}. If the time length of each Arnoldi iteration is defined as being equal to one time unit, this growth rate and this frequency correspond to the two dominant (conjugate) eigenvalues 
\begin{equation}
\lambda_{1,2} = 1.13857e^{{\pm}0.73944 \textbf{i}}.
\label{LeadingEV}
\end{equation}

As $|\lambda_{1,2}|>1$, the flow is unstable and the corresponding
instabilities exponentially grow through time. However the SFD method
was able to stabilise the flow and allowed it to converge towards its
steady-state.

We now verify that if the SFD method is able to stabilize the most
unstable eigenmode of a flow problem then the method will converge
towards the steady flow. In other words, if $\lambda_D$ is the
dominant unstable (\textit{i. e.} $|\lambda_D| > 1$) eigenvalue of a
flow problem. If the parameters $\chi_D$ and $\Delta_D$ enable to
reach the steady-state of the scalar problem $u^{n+1} = \lambda_D
u^n$. Then we want to know if the same parameters will also enable the
SFD method to reach the steady-state of the flow problem.
\begin{figure}
\begin{center}
\includegraphics[width=8.0cm]{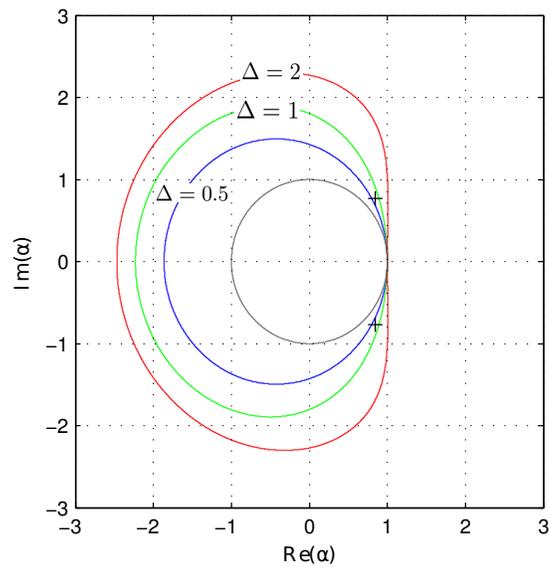} 
\caption{\label{StabContours} Contours of the stability regions of
  (\ref{1D_SplittingSFD}) for $\chi = 1$ and various $\Delta$. The
  central circle represents the boundary of the unit disc. The two
  black crosses indicate the position of the dominant unstable
  eigenvalues (\ref{LeadingEV}) of the cylinder flow at Re = 100.}
\end{center}
\end{figure}

\begin{figure}
\begin{center}
\subfigure[$ $]{\label{ConvHist} \includegraphics[width=8.0cm]{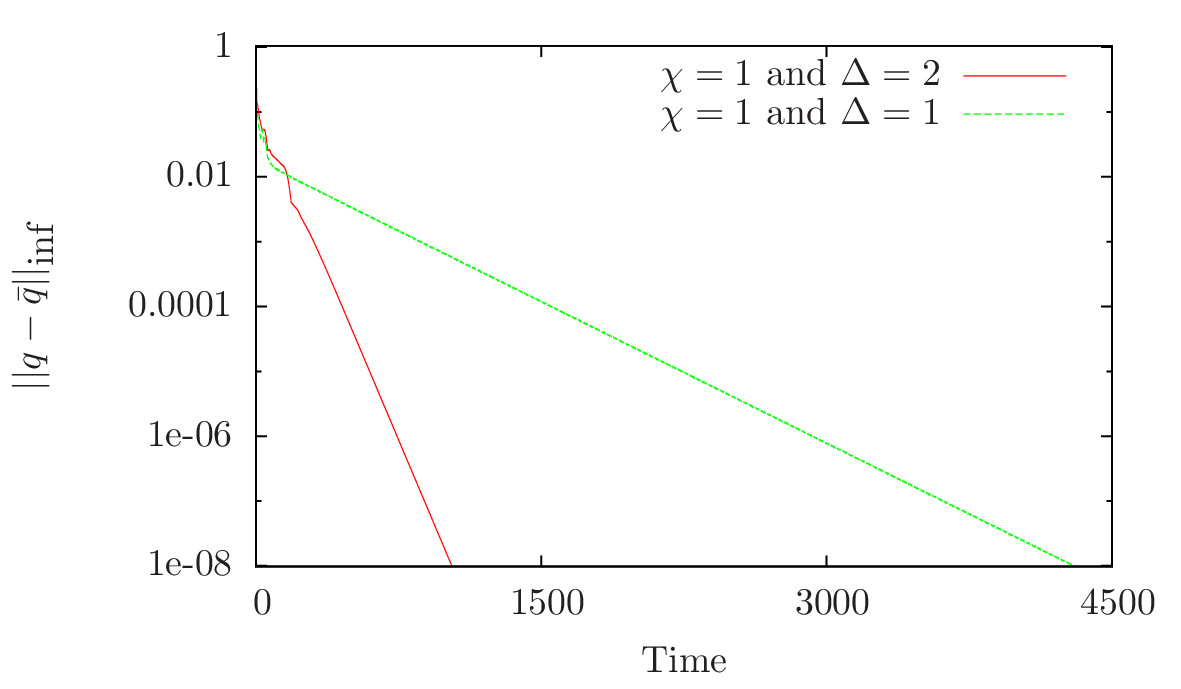}}
\subfigure[$ $]{\label{DivHist} \includegraphics[width=8.0cm]{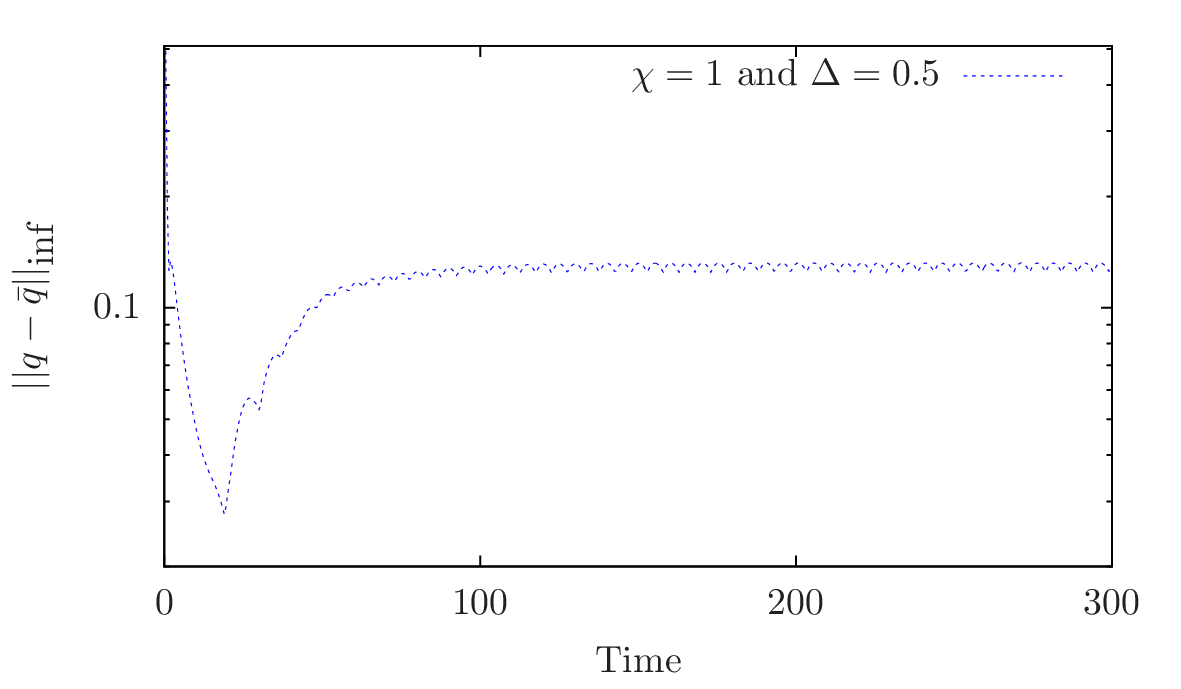}}
\caption{\label{ConvDivHistory} Time evolution of $||q -
  \bar{q}||_{\text{inf}}$ for parameters which allow the encapsulated
  SFD method to converge towards the steady-state \subref{ConvHist};
  and for parameters which do not \subref{DivHist}. The cases presented here have been computed with the time-step $\Delta t = 0.01$.}
\end{center}
\end{figure}

Figure \ref{StabContours} shows the contours of several stability
regions of (\ref{1D_SplittingSFD}), superposed with the position of
the dominant (conjugate) eigenvalues (\ref{LeadingEV}) of the cylinder
flow at $Re=100$. (Note that the stability regions of $\chi = 1$,
$\Delta = 0.5$ and $\chi = 1$, $\Delta = 2$ are the ones shown on
Figure \ref{StabRegions}\subref{Delta05} and
\ref{StabRegions}\subref{Delta2}). We notice that for $\chi = 1$ and
$\Delta = 1$ and also for $\chi = 1$ and $\Delta = 2$, the unstable
eigenvalues (\ref{LeadingEV}) are situated inside the stability region
of (\ref{1D_SplittingSFD}). These parameter couples have been used to
apply the SFD method to the Navier-Stokes solver and with both, the
steady solution was obtained. However the computational time required
to reach convergence strongly depends $\chi$ and $\Delta$. Figure
\ref{ConvDivHistory}\subref{ConvHist} compares the number of
iterations computed by the Navier-Stokes solver before obtaining the
steady-state. With $\chi = 1$ and $\Delta = 1$ the SFD method needs
about 4 times as many iterations to converge than with $\chi = 1$
and $\Delta = 2$.  Increasing the filter width does not always
decrease the computation time of the method. Indeed when $\Delta$
becomes too large, we obtain the expected limiting behaviour.

\begin{figure}
\begin{center}
\includegraphics[width=8.5cm]{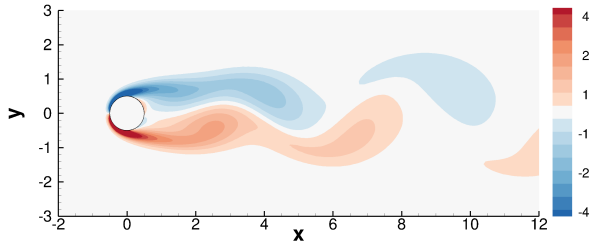}
\caption{\label{asdf}Vorticity of the partially controlled flow
  cylinder flow at $Re=100$. Snapshot obtained with SFD parameters
  which do not allow convergence towards the steady-state.}
\end{center}
\end{figure}

For $\chi = 1$ and $\Delta = 0.5$, the unstable eigenvalues
(\ref{LeadingEV}) are situated outside the stability region of
(\ref{1D_SplittingSFD}). When this parameter couple is used to apply
the SFD method to the cylinder flow, the steady-state can not be
obtained. Figure \ref{asdf} presents the outcome of the method with
these parameters. This flow is not steady but the oscillations are
attenuated in comparison with the flow presented in Figure
\ref{DiplayResultX1Delta2}\subref{UncontrolledFlow}. Figure
\ref{ConvDivHistory}\subref{DivHist} shows that when the SFD method is
not converging towards a steady-state, $||q - \bar{q}||_{\text{inf}}$
does not decrease but it oscillates around a fixed value.

In summary we can say that a relationship can be drawn
between the convergence (or not) of the SFD method applied to
(\ref{1DEquation}) with $\alpha = \lambda_D$ and the ability of the
SFD to drive the flow problem towards its steady-state. Note that the
problem studied here has only two (conjugate) unstable eigenmodes. 

We substituted $\alpha=\lambda_1$ and $\alpha=\lambda_2$ into our one
dimensional eigenvalue analysis, and numerically optimised $\chi$ and
$\Delta$ to minimise the maximum of the SFD growth rates for the two
eigenvalues. We obtained optimum parameters $\chi_{\text{opt}} \simeq
0.4391$ and $\Delta_{\text{opt}} \simeq 3.1974$. We observed that
these parameters gave the fastest SFD convergence when applied to the
cylinder flow (about 20\% faster than for $\chi = 1$ and $\Delta =
2$).

\section{Conclusion}

An alternative formulation of the SFD method, which enables us to use an
already existing time-stepping code as a "black box", is
presented. 
This method can be easily implemented as a wrapper function.
The convergence towards the unstable steady-state of the two-dimensional incompressible flow past
a cylinder at $Re=100$ is achieved without the use of a continuation
method, and the result matches with that of Barkley
\cite{BarkleyMeanVsSteadyCylinderFlow}.

The stability of the method relies on the oscillatory growth of the
problem studied. Indeed a problem which has a pure exponential growth
corresponds to a case where the dominant unstable eigenvalue would be
real. We have observed that the SFD method is not able to find the
steady-state of such cases (\textit{e. g.} wall confined jets
\cite{sherwin2005three}).

If the problem has unstable eigenvalues with an imaginary part, the
convergence towards the steady-state relies upon an appropriate choice
of the parameters $\chi$ and $\Delta$.  The knowledge of the dominant
eigenvalue $\lambda_D$ of the system allows us to select suitable
parameters through the analysis of the stability of the SFD method
applied to $u^{n+1} = \lambda_D u^n$. However most of the time the
dominant eigenvalue of a challenging flow problem is not known
\textit{a priori}. This suggests to investigate the possibility of
coupling the SFD method with an Arnoldi method which would evaluate
the dominant eigenvalue using a "partial" steady-state for base flow
(\textit{e. g.} the solution when $|| q^n - \bar{q}^n ||_{\text{inf}}
\simeq 10^{-2}$). The idea would be to obtain an approximation of the
dominant eigenvalue in order to be able to choose appropriate $\chi$
and $\Delta$.

\paragraph{Acknowledgments \\} 
The authors would like to thank the Seventh Framework Programme of the
European Commission for their support to the ANADE project (Advances
in Numerical and Analytical tools for DEtached flow prediction) under
grant contract PITN-GA-289428.

\end{document}